\documentclass[
    prd, twocolumn,
    superscriptaddress, nofootinbib, amsmath, amssymb,
    aps, floatfix
]{revtex4-1}

\usepackage{graphicx,xcolor,setspace}
\usepackage[
    colorlinks=true,
    linkcolor=blue,
    urlcolor=blue,
    citecolor=blue
]{hyperref}

\usepackage[capitalise]{cleveref}
\usepackage{siunitx}
\usepackage{color,epsfig}
\usepackage{ifpdf}
\usepackage{amsmath}
\usepackage{bm}
\usepackage[english]{babel}
\usepackage{amsfonts}%
\usepackage{amssymb}
\usepackage{braket}
\usepackage{hyperref}
\usepackage{enumerate}
\usepackage{cancel}
\usepackage{multirow}
\usepackage[compat=1.0.0]{tikz-feynman}

\usepackage{ulem}
\usepackage{array}
\usepackage{ulem,fancyvrb}
\usepackage{xcolor}

\definecolor{nicered}{rgb}{0.7,0.1,0.1}
\definecolor{nicegreen}{rgb}{0.1,0.5,0.1}
\hypersetup{colorlinks,citecolor= nicegreen,linkcolor= nicered}

\usepackage{dcolumn}
\usepackage{bm}
\usepackage{slashed}

\def\comments{true}

\ifx\comments\undefined
	\newcommand{\comment}[1]{}
\else
	\newcommand{\comment}[1]{#1}
\fi

\definecolor{maroon}{cmyk}{0,0.87,0.68,0.32}

\begin{document}

\title{Probing Supersymmetry Breaking Scale with Atomic Clocks}

\author{Victor V. Flambaum}
\email{v.flambaum@unsw.edu.au}
\affiliation{School of Physics, University of New South Wales, Sydney 2052, Australia}

\author{Xuewen Liu}
\email{xuewenliu@ytu.edu.cn}
\affiliation{Department of Physics, Yantai University, Yantai 264005, China}

\author{Igor Samsonov}
\email{igor.samsonov@unsw.edu.au}
\affiliation{School of Physics, University of New South Wales, Sydney 2052, Australia}

\author{Lei Wu}
\email{leiwu@njnu.edu.cn}
\affiliation{Department of Physics and Institute of Theoretical Physics, Nanjing Normal University, Nanjing, 210023, China}

\author{Bin Zhu}
\email{zhubin@mail.nankai.edu.cn}
\affiliation{Department of Physics, Yantai University, Yantai 264005, China}

\begin{abstract}
The supersymmetry (SUSY) breaking mechanism generally predicts the existence of the sgoldstinos, which can play the role of wave-like dark matter. Due to the ubiquitous coupling to the electromagnetic fields, the light scalar sgoldstino dark matter can lead to the variance of the fine-structure constant. With the precise atomic clock data, we find the SUSY breaking scale $\sqrt{F}$ can be probed up to the GUT scale in the sgoldstino mass range of $10^{-22}$ eV $<m_\phi <$ $4\times10^{-7}$ eV. 
\end{abstract}

\maketitle

\section{Introduction}

Supersymmetry (SUSY) is a postulated space-time symmetry between fermions and bosons. As one of the leading new physics models, SUSY has many impressive successes, such as the gauge coupling unification~\cite{Dimopoulos:1981yj} and the dark matter candidates~\cite{Goldberg:1983nd, Ellis:1983ew}. To connect with the physics at the weak scale, SUSY must be broken. Determining the SUSY breaking scale $\sqrt{F}$ will shed light on the new physics beyond the Standard Model.

The standard SUSY breaking approach is introducing a hidden sector, in which SUSY is spontaneously broken. Then, SUSY breaking effect is transmitted into the visible sector so that the masses of the supersymmetric particles (sparticles) depend on the SUSY breaking scale~\cite{Nilles:1983ge,Giudice:1998bp,Giudice:1998xp}. Direct search for sparticles at colliders provides one way to measure the fundamental SUSY breaking scale. The non-observation of the sparticles and the discovery of the Higgs boson~\cite{ATLAS:2012yve,CMS:2012qbp}, however, strongly indicate that the SUSY breaking scale is much higher than the weak scale. Besides, the current collider bounds on $\sqrt{F}$ are highly sensitive to the choices of the mediation scenarios, such as the gauge mediation, the gravity mediation, and the anomaly mediation, whereby any deviation from the three cases affects the reliability and efficiency. Therefore, finding an alternative model-independent way to probe the SUSY breaking scale is very crucial.

In models with the spontaneous SUSY breaking, the superpartners of a Goldstone fermion (goldstino), pseudoscalar $a$ and scalar $\phi$ goldstinos (sgoldstinos), must exist~\cite{Intriligator:2007cp}. After SUSY breaking, the goldstino is eaten and becomes the longitudinal component of the gravitino. Unlike the goldstino, the sgoldstinos are associated with non-compact flat directions in the tree-level scalar potential, called pseudo-moduli. Their masses are arbitrary, and can be of the order of the light gravitino mass or even substantially lighter~\cite{Brignole:1998uu}. The higher order terms in the Kahler potential may make the pseudo-moduli massive. However, as long as these terms are sufficiently tiny, the sgoldstinos can remain very light. Besides, the sgoldstinos naturally have couplings with electromagnetic fields, which are inversely proportional to the SUSY breaking scale~\cite{Perazzi:2000id}. As such, the searches for the light sgoldstinos via electromagnetic interactions can provide an alternative way of measuring the SUSY breaking scale. Among various existing constraints, the strongest bound comes from the astrophysical observation of the supernova $X$-ray, which excludes the SUSY breaking scale up to $\sim 10^{7}$ GeV through.

In this work, we propose to probe the SUSY breaking scale from the ultra-light CP-even sgoldstino field $\phi$ with the precision frequency techniques. In the early universe, the sgoldstino can be produced from a misalignment mechanism~\cite{Linde:1987bx,Preskill:1982cy}. Due to its interaction with the electromagnetic fields, it will lead to the variation of the fundamental couplings, such as the fine-structure constant~\cite{Arvanitaki:2014faa, Stadnik:2014tta, Stadnik:2016zkf}, and thus change the atomic transition frequency. We derive the oscillating-in-time shifts to the fine structure constant in high-scale SUSY. It is the first time that the atomic clock experiments place upper bounds on the SUSY breaking scale, which can be probed up to $\sqrt{F} \sim {\cal O}(10^{15})$ GeV in the sgoldstino mass range of $10^{-22}$ eV $<m_\phi <$ $4\times10^{-7}$ eV. Our result extends the bounds on SUSY break scale from other existing experiments.

\section{Ultra-light Sgoldstino DM}
In a realistic model, SUSY must be broken and predict the existence of sgoldstinos. Such pseudo-moduli fields appearing in the hidden sector are massless at tree level. However, due to the quantum corrections, these fields may obtain a tiny mass. We consider a $N=1$ globally supersymmetric model with two chiral superfields, $M$ and $\Phi$, where $M$ stands for the MSSM sector and $\Phi=\phi+i a+\theta \psi+\theta^2 F_{\Phi}$ comprises the goldstino and sgoldstinos. Then, the most general Kahler potential up to higher order derivatives becomes,
\begin{equation}
    K=M\overline{M}+\Phi\overline{\Phi}-\frac{M^2\overline{M}^2}{4\Lambda_M^2}-\frac{\Phi^2\overline{\Phi}^2}{4\Lambda_{\Phi}^2}-\frac{M\overline{M}\Phi\overline{\Phi}}{\Lambda^2}
\end{equation}
The mass scales $\Lambda$, $\Lambda_{M}$ and $\Lambda_{\Phi}$ determine the higher-order terms and are treated as the independent parameters. The superpotential is responsible for SUSY breaking, which remains the same as the tree-level result, $W=f\Phi$. It is the spurion analysis without relying on any assumptions on SUSY breaking sector. Computing the F-flatness condition, we obtain the SUSY breaking minima, 
\begin{equation}
    F_{\Phi}=f,\quad F_{M}=0.
\end{equation}
Then, SUSY is spontaneously broken with the vacuum energy $V=f^2$. As a result, the scalar potential around the minima can be written as
\begin{equation}
    V=f^2+m_{\phi}^2 (\phi^2+a^2)+m_{M}^2 M^2+\ldots
\end{equation}
with the sfermion mass $m_M$ and sgoldstino mass $m_{\phi}$ to be,
\begin{equation}
    m_M=\frac{f}{\Lambda},\quad m_{\phi}=\frac{f}{\Lambda_{\Phi}}.
\end{equation}
The mass hierarchy between sgoldstino and MSSM particles is thus fixed by the ratio $\Lambda_{\Phi}/\Lambda$. For a relevant ratio, the sgoldstino mass can be substantially lighter than MSSM particles. However, such a relationship suffers from the quantum correction to the effective Kahler potential. Due to the non-renormalization theorem~\cite{Seiberg:1993vc}, the quantum corrections behave as wave-function renormalization of superfields~\cite{Brignole:1998uu}, 
\begin{equation}
\begin{aligned}
\hat{M}&=\left[1-\frac{1}{2} \frac{\Lambda_{0}^{2}}{16 \pi^{2}}\left(\frac{1}{\Lambda_{M}^{2}}+\frac{1}{\Lambda^{2}}\right)\right] M\\
\hat{\Phi}&=\left[1-\frac{1}{2} \frac{\Lambda_{0}^{2}}{16 \pi^{2}}\left(\frac{1}{\Lambda^{2}}+\frac{1}{\Lambda_{\Phi}^{2}}\right)\right] \Phi\\
\hat{f}&=\left[1+\frac{1}{2} \frac{\Lambda_{0}^{2}}{16 \pi^{2}}\left(\frac{1}{\Lambda^{2}}+\frac{1}{\Lambda_{\Phi}^{2}}\right)\right]f\\
\frac{1}{\hat{\Lambda}_{M}^{2}}&=\frac{1}{\Lambda_{M}^{2}}+\frac{\Lambda_{0}^{2}}{16 \pi^{2}}\left(\frac{4}{\Lambda_{M}^{4}}+\frac{2}{\Lambda_{M}^{2} \Lambda^{2}}+\frac{2}{ \Lambda^{4}}\right)\\
\frac{1}{\hat{\Lambda}^{2}}&=\frac{1}{\Lambda^{2}}+\frac{\Lambda_{0}^{2}}{16 \pi^{2}}\left(\frac{3}{\Lambda^{4}}+\frac{2}{\Lambda_{M}^{2} \Lambda^{2}}+\frac{2}{\Lambda^{2} \Lambda_{\Phi}^{2}}\right)\\
\frac{1}{\hat{\Lambda}_{\Phi}^{2}}&=\frac{1}{\Lambda_{\Phi}^{2}}+\frac{\Lambda_{0}^{2}}{16 \pi^{2}}\left(\frac{4}{\Lambda_{\Phi}^{4}}+\frac{2}{\Lambda_{\Phi}^{2} \Lambda^{2}}+\frac{2}{\Lambda^{4}}\right)
\end{aligned}
\end{equation}
Keep track that the renormalized scalar masses arise immediately from the formulas, $\hat{m}_{M}=\hat{f}/\hat{\Lambda},\quad \hat{m}_{\Phi}=\hat{f}/\hat{\Lambda}_{\Phi}$.
The natural estimate is to set the cutoff $\Lambda_{0}$ comparable with the maximal energy scale $\Lambda_{\Phi}$. Then the order of estimate becomes,
\begin{equation}
\hat{f}\sim \left(\frac{\Lambda_{\Phi}^2}{\Lambda^2}\right)f,\quad \frac{1}{\hat{\Lambda}}\sim \left(\frac{\Lambda_{\Phi}}{\Lambda}\right)\frac{1}{\Lambda},
\quad \frac{1}{\hat{\Lambda}_{\Phi}}\sim \frac{{\Lambda}_{\Phi}}{\Lambda}\frac{1}{\Lambda}    
\end{equation}
Generally, the ratio between the MSSM particles and sgoldstino becomes $\mathcal{O}(1)$. To realize an ultra-light sgoldstino, the $\Lambda_{\Phi}$ should stabilize even after quantum correction, which requires a cancellation between the loop corrections. The negative contribution comes from the higher order terms in Kahler potential $\delta K=M\bar{M}\Phi^2\bar{\Phi}^2/4\Lambda_{\Phi M}^4$ so that the loop correction becomes,
\begin{equation}
    \hat{\Lambda}_{\Phi}\sim \frac{1}{\Lambda_{\Phi}}+\frac{\Lambda_0^2}{16\pi^2}\left(\frac{1}{\Lambda^4}-\frac{1}{\Lambda_{\Phi M}^4}\right)
\end{equation}
Once the $\Lambda_{\Phi M}$ equals $\Lambda$, the quantum correction can be negligible. Then we can obtain an ultra-light sgoldstino. We should point out that the significant fine-tuning, which is likewise the consequence of high-scale SUSY, comes at the expense of ultra-light sgoldstino.

In the early universe, such ultra-light sgoldstino can be produced through the misalignment mechanism. Due to its enormous number density, the cosmological evolution of its initial amplitude is related to the equation of motion,
\begin{equation}
\ddot{\phi}+3 H \dot{\phi}+V^{\prime}(\phi)=0.
\end{equation}
The sgoldstino and its energy density will become frozen when the Hubble scale exceeds the derivative of the scalar potential over the initial condition $\partial V(\phi_0)/\phi_0$ significantly. The sgoldstino harmonically oscillates as the Hubble scale falls below the critical point, which will behave like cold dark matter with amplitude $\phi_0=\sqrt{2\rho_{\phi}}/m_{\phi}$ and frequency $\omega \approx m_\phi$,
\begin{equation}
\label{eq7}
    \phi = \phi_0 \cos\omega t, 
\end{equation}
where $\rho_\phi$ is the sgoldstino energy density. $\rho_\phi$ can be related with the current dark matter density via $\rho_{\phi}=f_{\phi}\rho_{\mathrm{DM}}$, where $\rho_{\mathrm{DM}}=0.3\mathrm{GeV}/\mathrm{cm}^3$ and $f_\phi$ is the fraction of sgoldstino DM. Since this wave-like sgoldstino DM could couple with the electromagnetic fields, it will result in an effective time fluctuation of fundamental constants.

\section{Bounds on SUSY Breaking Scale from Atomic Clocks}
\label{SecClock}

The interaction of the sgoldstinos with the vector multiplet is given by
\begin{equation}
\mathcal{L}=\frac{c}{\Lambda} \operatorname{Re} \int d^{2} \theta\, \Phi W^{\alpha} W_{\alpha}\,,
\label{eqn:eft}
\end{equation}
where $c$ is a Wilson coefficient after integrating out the heavy messenger fields and $\Lambda$ corresponds to the UV scale. $\Phi=\varphi + \theta^2 F$ is a spurious superfield which comprises the SUSY breaking parameter $F$ and the complex scalar sgoldstino $\varphi = \phi + ia$. $W_{\alpha}$ is the superfield strength of the vector multiplet in the notation of Wess and Bagger~\cite{Wess:1992cp},
\begin{equation}
W_{\alpha}=-i \lambda_{\alpha}+\theta_{\alpha} D-\frac{i}{2}\left(\sigma^{\mu} \tilde{\sigma}^{\nu} \theta\right)_{\alpha} F_{\mu \nu}+\theta^{2}\left(\sigma^{\mu} \partial_{\mu} \bar{\lambda}\right)\,.
\end{equation}
Here $F_{\mu\nu}$ is the gauge field strength, $\lambda_{\alpha}$ is the gaugino field and $D$ is the auxiliary field. With the rule $\int d^{2} \theta\, \theta^{2}=1$, we can integrate out the Grassmann variables in Eq.~\ref{eqn:eft}, and then obtain the Lagrangian as, 
\begin{eqnarray}
\mathcal{L}&=& \frac{c}{2 \Lambda}\Big\{F \lambda^{\alpha} \lambda_{\alpha}+\bar{F} \bar{\lambda}_{\dot{\alpha}} \bar{\lambda}^{\dot{\alpha}}+2 a \partial_{\mu}\left(\lambda \sigma^{\mu} \bar{\lambda}\right) \nonumber \\ &&  -4 i \phi \lambda \sigma^{\mu} \partial_{\mu} \bar{\lambda} -2 i \partial_{\mu} \phi\left(\lambda \sigma^{\mu} \bar{\lambda}\right) \nonumber \\ &&
-\phi F_{\mu \nu} F^{\mu \nu}-a F_{\mu \nu} \tilde{F}^{\mu \nu}+2 \phi D^{2}\Big\}\,,
\label{eqn:expand}
\end{eqnarray}
where $\tilde{F}^{\mu \nu}=\frac{1}{2} \varepsilon^{\mu \nu \rho \sigma} F_{\rho \sigma}$. The auxiliary field $D$ can be eliminated by taking $D=0$. For simplicity, we assume the SUSY breaking parameter $F$ is a real constant. Then, we redefine the coefficients in Eq.~\ref{eqn:expand} as,
\begin{equation}
\frac{c}{\Lambda}=\frac{M}{F}\,,
\end{equation}
where $M$ is interpreted as the soft gaugino mass from the SUSY breaking effect. The terms without the derivatives of the spinors in Eq.~\ref{eqn:expand} are given by
\begin{eqnarray}
\mathcal{L} &\supset& \frac{M}{2} \lambda^{\alpha} \lambda_{\alpha}+\frac{M}{2} \bar{\lambda}_{\dot{\alpha}} \bar{\lambda}^{\dot{\alpha}} \nonumber \\ &&-\frac{1}{2} \frac{M}{F} \phi F_{\mu \nu} F^{\mu \nu}-\frac{1}{2} \frac{M}{F} a F_{\mu \nu} \tilde{F}^{\mu \nu},
\label{eqn:lag}
\end{eqnarray}
where $\phi$ and $a$ are the CP-even and CP-odd sgoldstino components, respectively. It should be noted that the interaction between the sgoldstino and the gauge bosons is independent of any assumptions on mediation scenarios, which provides a robust way to probe SUSY breaking scale. Moreover, due to the large value of $F$, the terms proportional to $ M / F$ in Eq.~\ref{eqn:lag} are usually ignored. However, when the sgoldstinos are light enough, such interactions will affect the low-energy observables, such as the atomic spectroscopy and the electric dipole moments. In this study, we will focus on the relevant CP-even sgoldstino interactions,
\begin{equation}
\label{eq6}
{\cal L}_{\phi FF}=-\frac12\frac{M}{F} \phi F_{\mu \nu} F^{\mu \nu},
\end{equation}
which can lead to the oscillating-in-time shifts to the fine structure constant,
\begin{equation}
\alpha\rightarrow  \alpha\left(1+\frac{2 M_{\tilde{\gamma}}}{F}\phi_0 \cos(\omega t)\right).
\end{equation}
Here $M_{\tilde{\gamma}}$ stands for the photino mass that is the combination of the bino mass $M_1$ and the wino mass $M_2$ after the electroweak symmetry breaking, 
\begin{equation}
M_{\tilde{\gamma}}=M_1 \cos^2\theta_w+M_2\sin^2\theta_w,
\end{equation}
where $\theta_w$ is the weak mixing angle. Such a variation in the fine structure constant will yield the drifts of the frequencies of atomic clocks, which can be calculated by,
\begin{equation}
\frac{\delta(\omega_{ 1}/\omega_{ 2})}{\omega_{ 1}/\omega_{ 2}} = 
K_\alpha \frac{\delta\alpha}{\alpha} + K_{e} \frac{\delta\mu_e}{\mu_e} 
+ K_{q} \frac{\delta\mu_q}{\mu_q}.
\label{fratio}
\end{equation}
Here $K_{\alpha,e,q}$ are the sensitivity coefficients that are obtained from the atomic, nucleon and nuclear structure calculations~\cite{Flambaum:2008kr,Dzuba:1999zz,Dzuba:1998au,Flambaum:2004tm,Flambaum:2006ip}. $\mu_e = m_e/m_p$ is the electron-proton mass ratio, and $\mu_q = m_q/\Lambda_{\rm QCD}$ is the ratio of the quark mass to QCD scale. The variation in the electron and quark masses results from the interaction between sgoldstino and the chiral multiplets, whose contributions are proportional to the soft trilinear term $A_0$. For simplicity, we assume the trilinear coupling $A_0=0$ so that the corrections to the fermion masses can be neglected in our study. Besides, given the null results of LHC searches for sparticles, we use the high-scale SUSY as a benchmark model to show the bounds on the SUSY breaking scale explicitly. In our numerical calculations, we take a common mass $M_{S}$ for all sparticles in the calculations. Then, we can rewrite Eq.~\ref{fratio} as,
\begin{equation}
\label{clock_frequency_ratio_quad}
\frac{\delta(\omega_{ 1}/\omega_{ 2})}{\omega_{ 1}/\omega_{ 2}}
=K_{\alpha}
\frac{\sqrt{2f_{\phi} \rho_{\mathrm{DM}}}}{m_{\phi} } \frac{2 M_{S}}{F} \cos( m_{\phi} t) 
\end{equation}

\begin{figure}[htbp]
\centering
\includegraphics[width=0.45\textwidth]{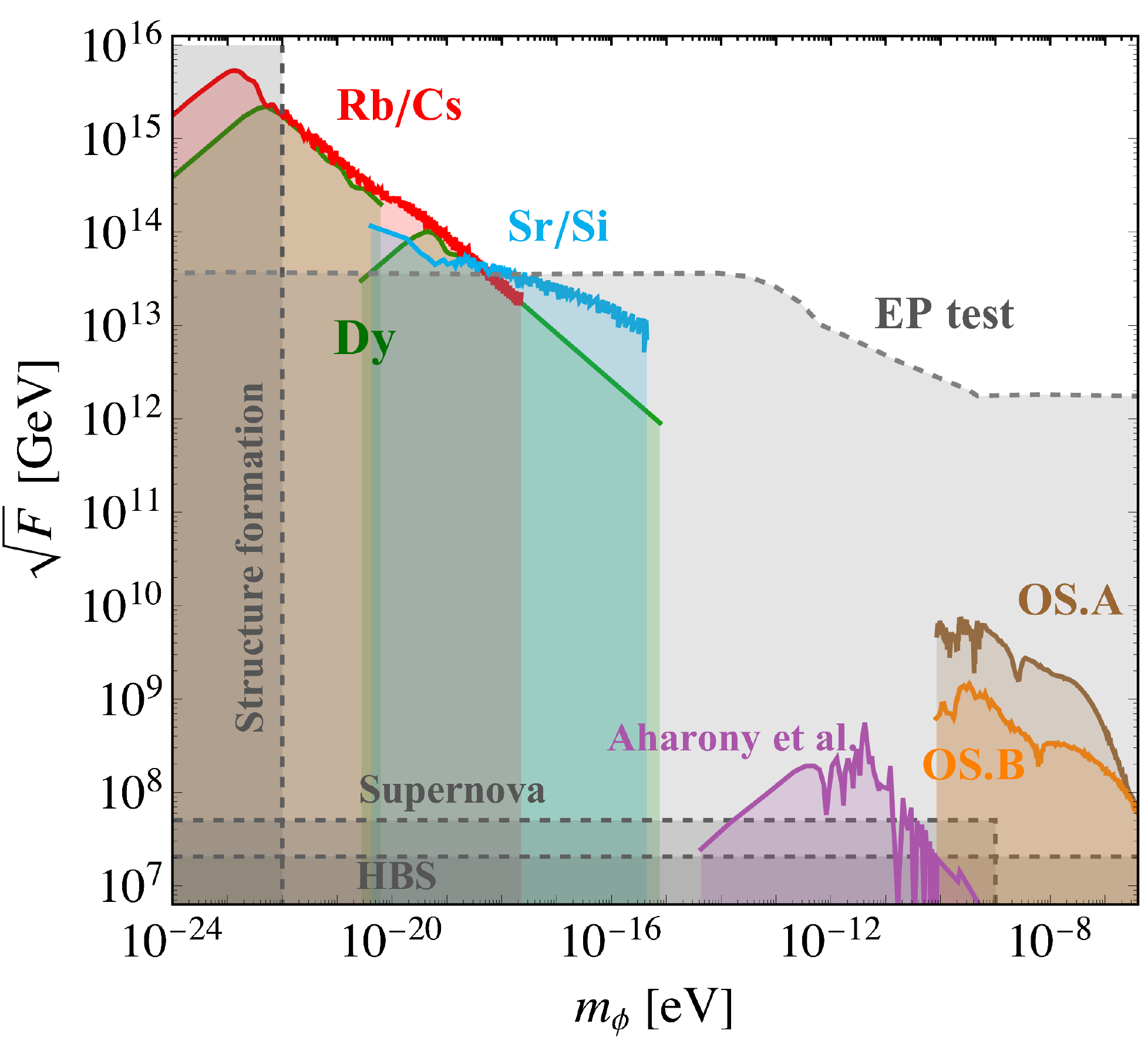}
\caption{Constraints on SUSY breaking scale $\sqrt{F}$ as functions of sgoldstino mass $m_{\phi}$. 
Lower limits on $\sqrt{F}$ from the Rb/Cs (red curve), Dy (green curves),  Sr/Si atom-cavity (blue curve) data are plotted. Purple bounds come from a dynamic decoupling experiment \cite{Aharony:2019iad}. The brown and orange curves represent the limits from searching for fast oscillations of fundamental constants by using atomic spectroscopy in cesium vapor\cite{Tretiak:2022ndx}.
The dashed-gray lines (regions)  stand for the excluded limits (regions)  arise from experiments testing the equivalence principle (EP test) \cite{Smith:1999cr,Schlamminger:2007ht,Berge:2017ovy,Hees:2018fpg}, the non-observation of $X$-ray excess from supernova  \cite{Brockway:1996yr,Gorbunov:2000th}, the helium-burning life-time of Horizontal Branch Stars (HBS) \cite{Raffelt:1996wa}, and the structure formation bound on the DM mass $m_\phi > 10^{-22}~\mathrm{eV}$~\cite{Hui:2016ltb}. Here we set $M_S=18$ TeV and $\tan \beta=20$ for producing the correct Higgs mass in high scale SUSY~\cite{Draper:2013oza}.}
\label{fig:constraints}
\end{figure}
In Fig.~\ref{fig:constraints}, we show the constraints on SUSY breaking scale $\sqrt{F}$ as a function of sgoldstino mass $m_{\phi}$. The hyperfine transition frequencies are measured in several processes~\cite{5422509,6174184,Gu_na_2014}, such as $|F=1,m_F=0\rangle \to |F=2,m_F=0\rangle$ 
for  $^{87}$Rb at $\approx$ 6.8~GHz and 
$|F=4,m_F=0\rangle \to |F=3,m_F=0\rangle$ 
for $^{133}$Cs at $\approx$ 9.2~GHz,  
at the FO2 dual fountain clock operating with caesium and rubidium atoms (${\rm ^{133} Cs/ ^{87} Rb}$) at LNE-SYRTE. Since there is no sinusoidal signature in the $^{87}$Rb/$^{133}$Cs atomic frequency ratio measurements, one can derive the 95\% C.L. upper bounds on the angular frequency $\omega$ via the power spectrum for each frequency $P(\omega)={N_o}/{4\sigma_o^2(\omega)}\mathcal A^2$ (where $N_o$ is the number of measurements, $\sigma_o^2(\omega)$ is their estimated variance~\cite{Hees:2016gop}, and the amplitude of oscillation $\mathcal A=K_{\alpha}\sqrt{2f_{\phi} \rho_{\mathrm{DM}}} 2 M_{S} /({m_{\phi} F})$). By using Eq.~\ref{clock_frequency_ratio_quad}, we obtain the lower limits on the SUSY breaking scale $\sqrt{F} \sim 1.8\times10^{13} \div 5\times10^{15}$ GeV in the sgoldstino mass range of $10^{-24} \lesssim m_\phi \lesssim 2\times10^{-18}$ eV (see the red lines in Fig.~\ref{fig:constraints}), where the sensitivity coefficient is taken as $K_\alpha=-0.49$ ~\cite{Flambaum:2008kr}. Note that there is also a lower bound on the DM mass being less than $10^{-22}$ eV from the structure formation~\cite{Hui:2016ltb}. Besides the Rb/Cs clock, the atomic spectroscopy measurements of dysprosium (Dy) \cite{VanTilburg:2015oza} are considered, as Dy has an electronic structure with two nearly degenerate levels whose energy splitting is very sensitive to changes in the fine structure constant. No signal has been observed in the two-year period data. It leads to the constraints on the SUSY breaking scale in the sgoldstino mass range of $10^{-24}  {\rm eV} \lesssim m_\phi \lesssim 10^{-15}$ eV (see the green lines in Fig.~\ref{fig:constraints}), which are weaker than those from Rb/Cs clock data. On the other hand, an atom-cavity frequency comparisons experiment has been performed~\cite{Kennedy:2020bac}, where they conduct the frequency comparisons between a strontium (Sr) optical lattice clock and a cryogenic crystalline silicon (Si) cavity. This provides stringent limits in the sgoldstino mass range of $10^{-21} {\rm eV}\lesssim m_\phi \lesssim 10^{-16}$ eV (see blue lines in Fig.~\ref{fig:constraints}), which are stronger than those from Dy data for $2\times 10^{-19} {\rm eV} \lesssim m_\phi \lesssim 10^{-16}$ eV. By utilizing the quantum information notion of dynamic decoupling, a light scalar DM within the mass range of $10^{-15}{\rm eV} \lesssim m_\phi \lesssim 4\times10^{-8}$ eV was not observed in an atomic optical transition~\cite{Aharony:2019iad}. The corresponding lower bounds on $\sqrt{F}$ are shown by the purple line  in Fig.~\ref{fig:constraints}. An optical-spectroscopy experiment~\cite{Tretiak:2022ndx}, searching for fast oscillations of fundamental constants, explores the the scalar mass range $8\times 10^{-11}{\rm eV} - 4\times 10^{-7} {\rm eV}$. Non-observation of the ultralight DM signature put bounds on the couplings to photons, we translate the limits from two different experimental setups on $\sqrt{F}$, which are depicted as OS.A and OS.B. We further show the constraints from the equivalence principle (EP) tests for comparison. One could see that the bounds of EP tests are stronger than that of the frequency measurements in $m_\phi\gtrsim 1.5\times 10^{-18} {\rm eV}$.

\section{Conclusion}
\label{sec:conclusion}
The sgoldstinos are predicted in models with spontaneous SUSY breaking. Searching for sgoldstinos is a robust way to probe the SUSY breaking scale. When the sgoldstino has a tiny mass and very weak interaction, it can play the role of wave-like dark matter. Due to the inevitable couplings with the electromagnetic fields, the sgoldstino dark matter can induce a periodic change of the fine structure constant. With the atomic clock data, we can probe the SUSY breaking scale up to the GUT scale in the sgoldstino mass range of $10^{-22}$ eV $<m_\phi <$ $4\times10^{-7}$ eV, which is far beyond the foreseen colliders and opens a new window of measuring SUSY breaking scale.

\begin{acknowledgments}
Lei Wu appreciates the helpful discussions with Fei Wang. This work was supported by the National Natural Science Foundation of China under the grants Nos. 11805161, and 12005180, by the Natural Science Foundation of Shandong Province under Grant No. ZR2020QA083, and by the Project of Shandong Province Higher Educational Science and Technology Program under Grants No. 2019KJJ007 and by the Australian Research Council Grants No. DP190100974 and DP200100150.

\end{acknowledgments}

\bibliography{refs}

\end{document}